\documentclass[aps,prc,groupedaddress,showpacs,twocolumn]{revtex4}

\usepackage{amsmath}
\usepackage{graphicx}

\begin{document}

\title{Transport model study of the $m_T$-scaling for $\Lambda$, K, and $\pi$ HBT-correlations}

\author {Qingfeng Li,$\, ^{1}$\footnote{E-mail address: liqf@fias.uni-frankfurt.de}
Marcus Bleicher,$\, ^{2}$ and Horst St\"{o}cker$\, ^{1,2,3}$}
\address{
1) Frankfurt Institute for Advanced Studies (FIAS), Johann Wolfgang Goethe-Universit\"{a}t, Max-von-Laue-Str.\ 1, D-60438 Frankfurt am Main, Germany\\
2) Institut f\"{u}r Theoretische Physik, Johann Wolfgang Goethe-Universit\"{a}t, Max-von-Laue-Str.\ 1, D-60438 Frankfurt am Main, Germany\\
3) Gesellschaft f\"ur Schwerionenforschung, Darmstadt (GSI), Germany \\
 }

%\date{\today}

\begin{abstract}
Based on the microscopic transport model UrQMD in which hadronic and
string degrees of freedom are employed, the HBT parameters in the
longitudinal co-moving system are investigated for charged pion and
kaon, and $\Lambda$ sources in heavy ion collisions (HICs) at SPS
and RHIC energies. In the Cascade mode, $R_O$ and the $R_L$ at high
SPS and RHIC energies do not follow the $m_T$-scaling, however,
after considering a soft equation of state with momentum dependence
(SM-EoS) for formed baryons and a density-dependent Skyrme-like
potential for ``pre-formed'' particles, the HBT radii of pions and
kaons and even those of $\Lambda$s with large transverse momenta
follow the $m_T$-scaling function $R=3/\sqrt{m_T}$ fairly well.
\end{abstract}


 \maketitle

In order to explore the properties of the fireball which is created
just after the collision of two energetic nuclei, the
Hanbury-Brown-Twiss interferometry (HBT) technique has been used and
developed in astro- and nuclear-particle physics for about half a
century.  It is well-known that the HBT technique can provide
important information about the spatio-temporal structure of the
particle emission source (the region of homogeneity). In
\cite{Rischke:1996em}, a non-trivial transition in the excitation
function of spatio-temporal characteristics of the source was
proposed. Meanwhile, the AGS, SPS and RHIC experiments (with
nucleon-nucleon center-of-mass energies $\sqrt{s_{NN}}$ from about
2.5 GeV up to 200 GeV for heavy ion collisions) have been
stimulating the HBT related investigations further into another
golden era. Although the experiments have discovered quite a few
exciting hints for a new phase of matter - maybe a quark gluon
plasma (QGP) (see, e.g., Refs.\ \cite{Adams:2005dq,Harris:2007zza}),
a plethora of puzzling phenomena also came out. Examples related to
HBT are,
\begin{enumerate}
\item the
discovered that there is no {\it obvious} peak/valley of the HBT
quantities over the whole beam energies from SIS, AGS, SPS, up to
RHIC as suggested in \cite{Rischke:1996em}.
 \item that hydro-dynamic as well as transport (cascade
mode) calculations show that the calculated ratio of the HBT radii
in outward and sideward directions is higher than the experimental
data, which is known as the HBT time-puzzle (``t-puzzle``)
\cite{Lisa:2005dd,Li:2006gp,Li:2006gb,Li:2007im}. For possible
solutions of the ''t-puzzle'' we refer to
\cite{Lin:2002gc,Cramer:2004ih,Humanic:2005ye,Pratt:2007pb,Li:2007yd,Romatschke:2007jx}.
 \item mainly due to the resonance decay contribution of the particle
production as well as the effects of strong and electromagnetic
forces, the correlation function of identical particles deviates
from a Gaussian type \cite{Wiedemann:1999qn,Lisa:2005dd}.

\end{enumerate}

 A ``direct'' study of the so-called ``non-Gaussian effect'' --- the
emission source image technique--- has been quickly improving in
recent years
\cite{Chung:2007yq,Danielewicz:2006hi,Lacey:2007rj,Afanasiev:2007kk}.
However, in order to theoretically investigate previous data and due
to the fact that it gives a leading-order approximation to the real
shape of the homogeneity region, the Gaussian parameterization is
still important. During the process of Gaussian parameterization, it
is known that on both experimental and theoretical sides, a proper
potential modification of the final state interaction (FSI) after
freeze-out should be analyzed before any work is done because of the
intrinsic physical characteristic of correlated pairs. For example,
charged pion-pion interferometry is only relatively weakly affected
by the Coulomb and nuclear potentials due to its small mass and
collision cross section, while for charged kaon-kaon case, the
Coulomb modification should be taken into account in analysis
\cite{Lisa:2005dd,Lin:2003iq}. For baryon-baryon correlation, a
proper nuclear potential modification has to be considered. In the
absence of FSI and from a hydro-dynamic point of view in which the
freeze-out of particles is flow-dominated, the HBT radii are
predicted to decrease with $1/\sqrt{m_T}$ ($m_T$-scaling), $m_T$ is
the transverse mass of the observed particle-pair, independent of
the particle species \cite{Csorgo:1995bi}. This prediction has been
probed by recent experiments with energies from AGS to RHIC. In
Ref.\ \cite{Bearden:2001sy}, a $m_T$-scaling expression
$3/\sqrt{m_T}$ has been suggested at $E_{lab}=158A$ GeV. It is also
supported by other experiments at RHIC energies (see references in,
e.g., \cite{Lisa:2005dd}). However, due to the lack of data and the
rather large experimental errors, a firm conclusion is still
awaiting. It is also urgent to perform microscopic transport-model
calculations to explore possible deviation from the ideal
$m_T$-scaling especially in order to understand `non-Gaussian''
effects as well as the HBT puzzles.

In this paper, we investigate the $m_T$ dependence of the HBT
parameters with three kinds of identical-particle correlations:
$\pi^--\pi^-$, $K^+-K^+$, and $\Lambda-\Lambda$. So far the most
majority of HBT investigations are on the $\pi-\pi$ pairs due to their
large yield. In order to obtain a cleaner signal with sensitivity to
earlier stages of HICs, kaon-kaon correlations have been
explored at several energies
\cite{Bearden:2001sy,Afanasiev:2002fv,Bekele:2004ci,Lisa:2005dd,Abelev:2006gu}.
Besides mesons, in order to map out a widespread $m_T$-dependence,
the $\Lambda-\Lambda$ baryonic correlation is the next
convenient choice. It is an identical non-charged-particle
correlation so that a similarly Gaussian shape as for pions can be expected if
the nuclear potential is not considered in FSI.

To explore the $m_T$-scaling, we employ the UrQMD model. In UrQMD,
the hadrons are represented by Gaussian wave packets in phase space.
After the Wood-Saxon initialization, the phase space of hadrons is
propagated according to Hamilton's equation of motion
\cite{Bass98,Bleicher99}, $ {\bf \dot{r}}_i=\frac{\partial
H}{\partial {\bf p}_i}$ and ${\bf \dot{p}}_i=-\frac{\partial
H}{\partial {\bf r}_i}$. Here ${\bf {r}}_i$ and ${\bf {p}}_i$ are
the coordinate and the momentum of hadron $i$. The Hamiltonian $H$
consists of the kinetic energy $T$ and the effective interaction
potential energy $U$, $H=T+U$. The two-body Coulomb potential is
considered for formed charged particles. Recently, a soft equation
of state with momentum dependence (SM-EoS) for formed hadrons and a
density dependent Skyrme-like term for ``pre-formed'' hadrons from
string fragmentation have been supplied into the UrQMD transport
model, please see details in \cite{Li:2007yd,lqf20063}. For
observables such as the nuclear stopping, the elliptic flow and the
HBT parameters of pions visible improvement towards the data has
been observed if these potentials are included. Especially, the HBT
``t-puzzle'' for pions can be consistently solved. In this paper, we
continue this topic and further calculate the HBT parameters of kaon
and $\Lambda$ sources.

Three cases of experimental data at mid-rapidity are compared with
our calculations and the phase space cuts are same as those listed
in \cite{Li:2007im}, the three systems under investigation are: (1)
central Pb+Pb collisions at the SPS beam energy $E_{lab}=20 A$ GeV
(dubbed ''E20''), (2) central Pb+Pb collisions at the SPS beam
energy $E_{lab}=158 A$ GeV (''E158''), (3) central Au+Au collisions
at the top RHIC energy $\sqrt{s_{NN}}=200$ GeV (''s200''). For each
case about $25$ thousand central events are calculated. If not
stated otherwise, the UrQMD transport program stops at
$t_{cut}=200$fm$/c$, the residual long-lived unstable resonances are
forced not to decay after this final cut time. All particles with
their phase space coordinates at their respective freeze-out time
$t_f$ (last collisions) are put into the analyzing program to be
discussed below. When studying the time-evolution of the particle
correlation (shown in Fig.\ \ref{fig4}), we produce output at
$t_{cut}=5$, $10$, $15$, $20$, $25$, $50$, and $75$fm$/c$,
separately, while other parameters are not altered.

The ``correlation after-burner" (CRAB v3.0$\beta$) program
\cite{Pratt:1994uf,Pratthome} is then adopted for analyzing the
interactions of two particles after freeze-out with quantum
statistics and final state modifications so that one can construct
the HBT correlator to be compared with experimental data. In this
work, the strong interaction of pions and kaons is not
considered in FSI because of its negligible effect
\cite{Adams:2004yc}. However, we also noticed that the effect of
strong FSI influences the final HBT results of neutral $K_s^0$
source \cite{Abelev:2006gu}. For $\Lambda$s, the nuclear modification
is not considered in this work for simplicity. We will, however, briefly
discuss this issue in Fig.\ \ref{fig3}. Concerning the
contribution of the final state Coulomb interaction to the
correlator, it is checked for the charged pion-pion and kaon-kaon
cases. One billion pairs are performed at mid-rapidity of single
or two correlated particles as stated in \cite{Li:2007im} in each
CRAB analyzing run.

In the next step, we fit the correlator as a three-dimensional
Gaussian form under the Pratt convention (using ROOT
\cite{roothomepage} and the $\chi$-squared method), i.e., the
longitudinally co-moving system (LCMS). When the nuclear and Coulomb
modifications are not considered in the correlator, the fitting function
can be expressed in the standard way,
\begin{equation}
C(q_L,q_O,q_S)=1+\lambda
e^{-R_L^2q_L^2-R_O^2q_O^2-R_S^2q_S^2-2R_{OL}^2q_Oq_L}. \label{fit1}
\end{equation}
In Eq.~(\ref{fit1}), $\lambda$ is normally referred to as an
incoherence factor. It might be also affected by many other factors,
such as the contaminations, long-lived resonances, or the
details of the Coulomb modification in FSI. Thus, we regard it
as a free parameter.  $R_L$, $R_O$, and $R_S$ are the Pratt radii in
longitudinal, outward, and sideward directions, while the cross-term
$R_{OL}$ plays a role at large rapidities. $q_i$ is the pair
relative momentum $\mathbf{q}$
($\mathbf{q}=\mathbf{p}_1-\mathbf{p}_2$) in the $i$ direction.

If one considers the Coulomb effect in FSI for charged-particle
pairs, a Bowler-Sinyukov method, which has been used in STAR
experiments \cite{Adams:2004yc}, can be exploited in the fitting
process:

\begin{eqnarray}
&&C(q_L,q_O,q_S)=(1-\lambda)  + \nonumber \\
                            &&\lambda K_{coul}(q_{inv})
(1+e^{-R_L^2q_L^2-R_O^2q_O^2-R_S^2q_S^2-2R_{OL}^2q_Oq_L}),
\label{fit2}
\end{eqnarray}
where the $K_{coul}$ is the Coulomb correction factor and depends
only on $q_{inv}$ as same as experiments. The
$q_{inv}=\sqrt{\mathbf{q}^2-(\Delta E)^2}$ is the invariant relative
momentum, where $\Delta E=E_1-E_2$ is the energy difference of two
particles.

\begin{figure}
\includegraphics[angle=0,width=0.48\textwidth]{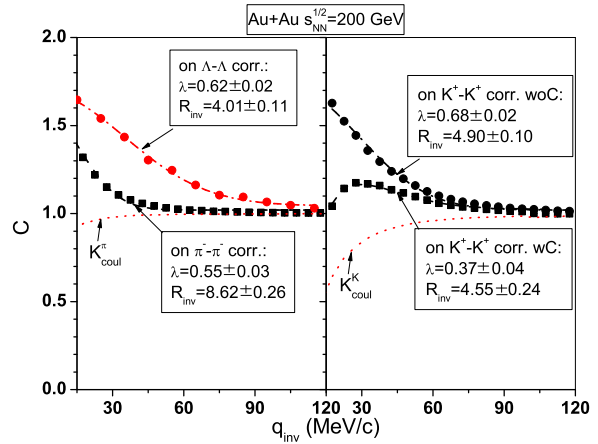}
\caption{One-dimensional invariant correlation function for pion,
$\Lambda$, and kaon sources for ``s200'' case. The potentials for
both ``pre-formed'' and formed particles are considered in
calculations. In the left plot: The squares and circles represent
the calculation results for pion and $\Lambda$ source without
modifications on FSI, separately. The dashed and dash-dotted lines
are the fitting results to them. In the right plot: The squares and
circles represent the calculation results for kaon source with and
without Coulomb modification, separately. The dashed and dash-dotted
lines are the fitting results to them. The Coulomb correction
factors of $\pi^--\pi^-$ and $K^+-K^+$ pairs ($K_{coul}^\pi$ and
$K_{coul}^K$) are also shown in both plots by thin dotted lines. }
\label{fig1}
\end{figure}

Fig.\ \ref{fig1} shows the one-dimensional invariant correlation
function for pion (left plot), $\Lambda$ (left plot), and kaon
(right plot) sources for central Au+Au reactions at
$\sqrt{s_{NN}}=200$ GeV. A transverse momentum $k_T$ cut is not
applied. The pairs are binned with $\Delta q_{inv}=5$MeV$/c$ for
pions and kaons while $\Delta q_{inv}=10$MeV$/c$ for $\Lambda$s due
to the small yield of $\Lambda$s in each event. The upper limit is
$q_{inv}=120$MeV$/c$. The pairs within $\mathbf{q}<5$MeV$/c$ are not
used for fitting due to the large errors from split and merged
tracks in the experiments. The left plot shows that the Gaussian
parameterization is still suitable for pion-pion and
$\Lambda$-$\Lambda$ correlators if strong and Coulomb modifications
are not considered in FSI. It was found that the Coulomb correction
factor influences the correlator of pions at
$q_{inv}\lesssim30$MeV$/c$ \cite{Wiedemann:1999qn,Adamova:2002wi}
which is supported by the present calculation as indicated by the
flat $K_{coul}^\pi$ (thin dotted line). It is also found that the
$K_{coul}^\pi$ factor in Eq.~(\ref{fit2}) reduces the final HBT
radii of pion source only slightly at $k_T\lesssim 100$MeV$/c$,
while the reduction of $\lambda$ is about $0.1$. A similar effect of
Coulomb modification on the HBT radii and the $\lambda$ factor was
also observed in previous calculations
\cite{Pratt:1986ev,Hardtke:1997cy}. The one-dimensional fitting
results ($R_{inv}$) for pions and $\Lambda$s show that the pion's
homogeneity length is about twice larger than for $\Lambda$'s, which
implies that more $\Lambda$s are emitted from the early stage. In
addition, the larger $\lambda$ factor indicates that $\Lambda$s are
less affected by the decay of resonances at late stage, although
some amount of $\Lambda$s are produced by the decay of $\Sigma^*$
resonances. Fig.\ \ref{fig1} (right) shows the $R_{inv}$ of the
$K^+$ source which lies between that of the pion and $\Lambda$
source, also the $\lambda$ value of the kaon source is largest,
which supports that kaons are produced earliest among the
investigated particles \cite{Bass98}. However, for kaons, the
Coulomb modification alters the correlator up to the large value of
$q_{inv}$, $\sim 60$MeV$/c$. Although a strong reduction of the
$R_{inv}$ is not seen, $\lambda$ is reduced by $\sim 0.3$. The
Coulomb corrected $\lambda$ values of pions and kaons match the data
much better \cite{Adler:2004rq,Adams:2004yc,Bekele:2004ci}. Due to
the fact that the Coulomb modification to the kaon-kaon correlation
is larger than for the pion-pion correlation, we adopt the
Bowler-Sinyukov fitting method expressed in Eq.~(\ref{fit2}) after
constructing the 3-dimensional correlator of kaon source with
Coulomb modifications.

\begin{figure}
\includegraphics[angle=0,width=0.48\textwidth]{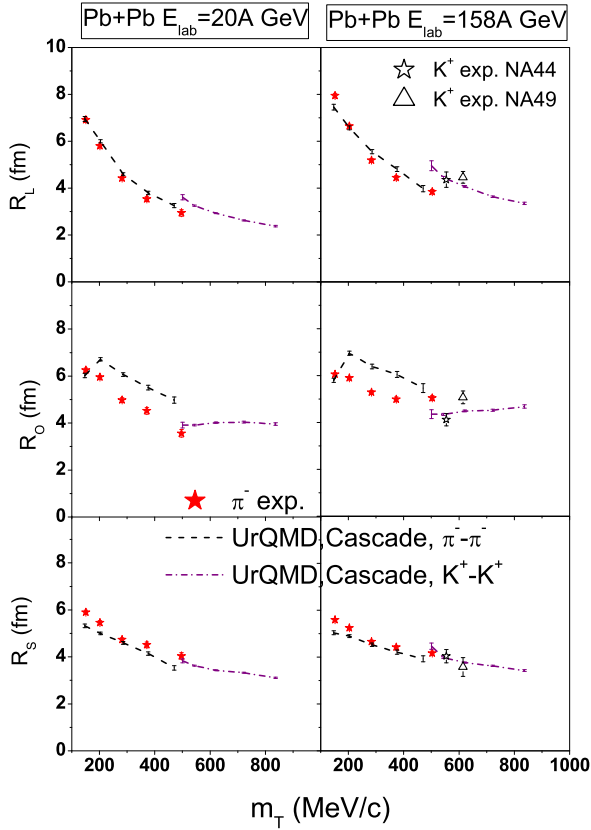}
\caption{$m_T$-dependence of the HBT radii of $\pi^-$ (dashed lines)
and $K^+$ (dash-dotted lines) sources in central $Pb+Pb$ collisions
for ``E20'' (left plots) and ``E158'' (right plots) cases (with the
cascade mode in UrQMD calculations). The experimental NA49 and NA44
data of pions (for ``E20'' and ``E158'') and kaons (for ``E158'')
are also shown by solid and open symbols, separately
\cite{Kniege:2006in,Bearden:2001sy,Afanasiev:2002fv}. } \label{fig2}
\end{figure}

\begin{figure}
\includegraphics[angle=0,width=0.48\textwidth]{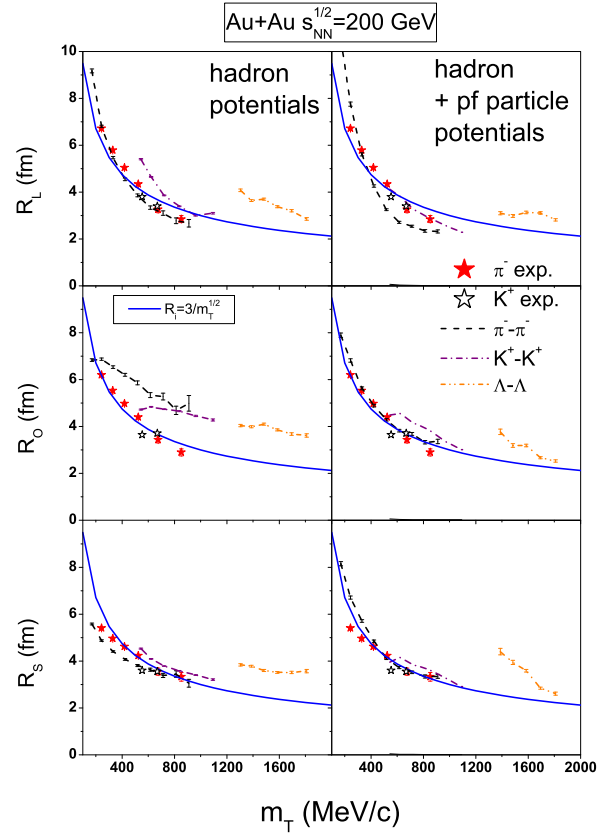}
\caption{$m_T$-dependence of the HBT radii of $\pi^-$ (dashed
lines), $K^+$ (dash-dotted lines), and $\Lambda$ (dash-dot-dotted
lines) sources in central $Au+Au$ collisions for ``s200'' case with
potentials only for formed baryons (left plots) and with potentials
for both formed baryons and ''pre-formed'' particles (right plots).
The experimental RHIC data of pions and kaons are shown by solid and
open stars separately
\cite{Adler:2004rq,Adams:2004yc,Bekele:2004ci}. The function
$R_i=3/\sqrt{m_T}$ is also shown by solid line in each plot.}
\label{fig3}
\end{figure}

\begin{figure}
\includegraphics[angle=0,width=0.48\textwidth]{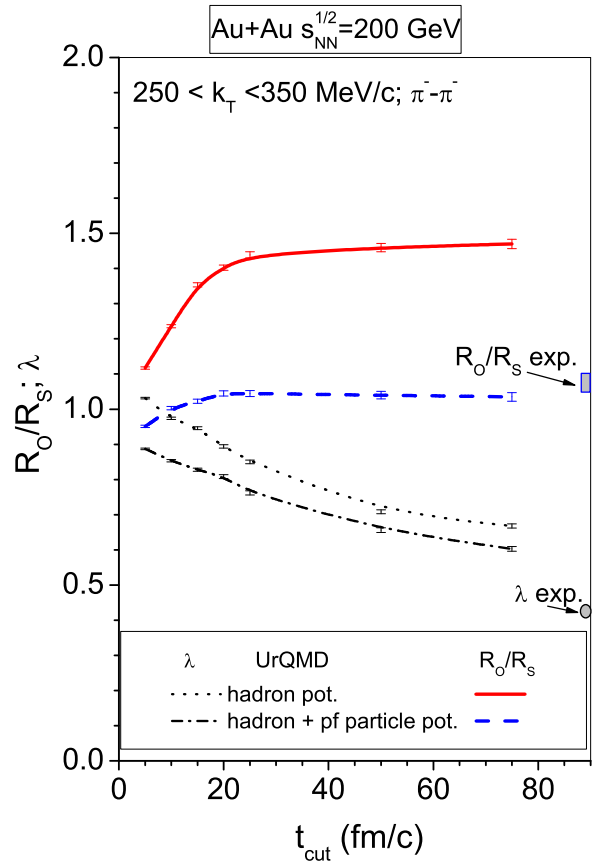}
\caption{Time evolution of the ratio $R_O/R_S$ and the $\lambda$
factor of pion source at $250<k_T<350$MeV$/c$, which are calculated
with potentials for formed baryons and with potentials for both
formed and ``pre-formed'' particles, are shown by different lines.
The $R_O/R_S$ and $\lambda$ experimental data are represented by a
square and and a circle at the right end of the plot. } \label{fig4}
\end{figure}

The HBT parameters can be as a function of $m_T$, where
$m_T=\sqrt{k_T^2+m^2}$ where $m$ is the pion, kaon, or $\Lambda$
mass and $\mathbf{k}_T=(\mathbf{p}_{1T}+\mathbf{p}_{2T})/2$ is the
transversal component of the average momentum $\mathbf{k}$ of two
particles, $\mathbf{k}=(\mathbf{p}_1+\mathbf{p}_2)/2$. Fig.\
\ref{fig2} depicts the calculated $m_T$-dependence of the HBT radii
$R_L$ (top plots), $R_O$ (middle plots), and $R_S$ (bottom plots) of
$\pi^-$ (dashed lines) and $K^+$ (dash-dotted lines) sources in
central $Pb+Pb$ collisions at $E_{lab}=20$A GeV (left plots) and at
$E_{lab}=158$A GeV (right plots). The cascade results are compared
with data (solid stars for pion data at both beam energies
\cite{Kniege:2006in}, open symbols for kaon data at 158 GeV from the
NA44 and the NA49 collaboration
\cite{Bearden:2001sy,Afanasiev:2002fv}). One should bear in mind
that the kaon data are obtained under slightly different physical
cuts from the pion case. For instance, in Ref.\
\cite{Afanasiev:2002fv}, the $5\%$ most central interactions are
selected for the central kaon data sample while the $7.2\%$ most
central interactions are selected for pions \cite{Kniege:2006in}. At
$E_{lab}=20A$ GeV, we observe $m_T$-scaling for pions and kaons in
$R_L$ and $R_S$, only the calculated $R_O$ of the pion source is
slightly larger than the data. It is also seen that the
$m_T$-dependence of $R_O$ of kaon source is weak. At $E_{lab}=158A$
GeV, the $m_T$-scaling of $R_S$ and $R_L$ still hold with similar
problems for $R_O$. However, one should note that the few currently
available kaon data can not sufficiently evaluate the performance of
the scaling.

Let us move on to see the $m_T$-dependence of HBT radii of pions
(dashed lines), kaons (dash-dotted lines), and $\Lambda$s
(dash-dot-dotted lines) at the top RHIC energy $\sqrt{s_{NN}}=200$
GeV, which is shown in Fig.\ \ref{fig3}. The pion data (solid stars)
are from \cite{Adler:2004rq,Adams:2004yc}, and the preliminary kaon
data (open stars) are from \cite{Bekele:2004ci}. The left plots show
the calculations with the SM-EoS for formed baryons, while the
calculations with potentials for both ``pre-formed'' and formed
particles are shown in the right plots. In each plot, the
$m_T$-scaling function $R_L=R_O=R_S=3/\sqrt{m_T}$ is also shown by
solid line. In Fig.\ \ref{fig3} (left), one can not observe
$m_T$-scaling in all HBT directions if the potentials are considered
only for formed baryons. Although the calculated $R_S$ of kaons stay
on the scaling line, those of $\Lambda$s are about $1.2$ fm above
the scaling line. Further, $R_L$ values of kaons and $\Lambda$s are
larger than the scaling results. In the $R_O$ direction the same is
true for the HBT radii of pions, kaons and $\Lambda$s. Let us now
consider potentials for ``pre-formed'' hadrons as discussed in
\cite{Li:2007yd}. Fig.\ \ref{fig3} (right) shows that the transverse
radii $R_O$ and $R_S$ of the pion source nicely follow the scaling
line. However, a steeper $m_T$-dependence of $R_L$ is seen which
might be due to the absence of a proper collision term for the
``pre-formed'' hadrons. Secondly, it is exciting to see that the
``pre-formed'' hadron potential leads to a smaller $R_L$ ($R_O$) of
kaon pairs as well so that the HBT radii of kaons also follow the
scaling line quite well. Thirdly, the results for $\Lambda$s also
approach towards the scaling line especially at large $k_T$ where
the effect of resonance decay is minor because of an early emission.
At small $k_T$, the decay of long-lived resonance $\Sigma$(1385)
into $\Lambda$s enlarges the HBT radii. Meanwhile, the absence of
the nuclear modification on FSI might also play a role, which
deserves a further investigation.

Based on discussions in Figs.\ \ref{fig1} - \ref{fig3}, we conclude
that interactions at the late stage (including the modifications in
FSI) indeed influences the HBT parameters. The idea of
''pre-formed'' hadron interactions does not only solve the HBT
''time ($R_O/R_S$)-puzzle'', but also improves the $m_T$-scaling
considerably. To elaborate further on this point, Fig.\ \ref{fig4}
shows the time evolution of the ratio $R_O/R_S$ and the incoherence
factor $\lambda$ of pion pairs at $250<k_T<350$MeV$/c$ for central
Au+Au reactions at $\sqrt{s_{NN}}=200$ GeV. Calculations with and
without ``pre-formed'' hadron potential are compared. The
corresponding experimental data of the $R_O/R_S$ and $\lambda$ are
indicated by a square and a circle at the right end of the plot.
Without ''pre-formed'' particle interactions, the $R_O/R_S$ ratio
increases rapidly up to $t\sim 25$fm$/c$ from $\sim 1.0$ to $\sim
1.5$. In contrast, considering the ``pre-formed'' hadron potential,
the ratio $R_O/R_S$ is nearly time independent (stays at its start
value of $R_O/R_S\sim 1\pm 0.05$). The effects of FSI continue to
play visible roles (although relatively weakly after $25$fm$/c$) on
the $\lambda$ factor, which is also consistent with the visible
suppressing effect of the Coulomb modification of FSI on the
$\lambda$ value from the analysis of Fig.\ \ref{fig1}. We also find
that the ``pre-formed'' hadron potential reduces the $\lambda$
further to approach the data.

For central collisions, the HBT radii can be (approximately)
analytically obtained under the assumptions of thermalization and
Gaussian-source shape and be expressed as
\cite{Herrmann:1994rr,Wiedemann:1999qn},

\begin{subequations}
\begin{align}
R_L^2 & =\langle (\widetilde{z}-\beta_L\widetilde{t})^2\rangle, \\
R_O^2 & =\langle (\widetilde{x}-\beta_T\widetilde{t})^2\rangle, \\
R_S^2 & =\langle \widetilde{y}^2 \rangle.
\end{align} \label{eqRlRoRs}
\end{subequations}
Here the space-time coordinates $\widetilde{x}$, $\widetilde{y}$,
$\widetilde{z}$, and $\widetilde{t}$ are relative distances to their
``effective source centers'' ($\overline{x}^\mu=\langle
x^\mu\rangle$): $\widetilde{x}^\mu=x^\mu-\overline{x}^\mu$. And
$\beta_L$ and $\beta_T$ are components of the velocity of particle
pair $\mathbf{\beta}$ ($\mathbf{\beta}=\mathbf{k}/k^0$,
$k^0=(E_1+E2)/2\approx \sqrt{m^2+\mathbf{k}^2}$ is the average
energy of two particles. Usually, the on-shell approximation is
used). Eq.~(\ref{eqRlRoRs}b) can also be expanded as
\begin{equation}
R_O^2=\langle\widetilde{x}^2\rangle+\langle\beta_T^2\widetilde{t}^2\rangle-2
\langle \beta_T\widetilde{x}\widetilde{t}\rangle. \label{eqRo}
\end{equation}
In central collisions, because of the longitudinal reflection
symmetry, $\langle\widetilde{x}^2\rangle\simeq
\langle\widetilde{y}^2\rangle$. Therefore, by comparing
Eq.~(\ref{eqRlRoRs}c) with Eq.~(\ref{eqRo}), it is clear that the
difference of $R_O$ and $R_S$ mainly comes from the relative
strength of the time-related term
$\langle\beta_T^2\widetilde{t}^2\rangle$ and the
$\widetilde{x}$-$\widetilde{t}$ correlation term $-2 \langle
\beta_T\widetilde{x}\widetilde{t}\rangle$. Thus, one obtains a clear
interpretation of the large reduction of the $R_O/R_S$ ratio (even
$R_O/R_S<1$ may happen), if ''pre-formed'' particle interactions are
taken into account by relating it to the stronger phase-space
correlation induced by the potentials.

\begin{figure}
\includegraphics[angle=0,width=0.48\textwidth]{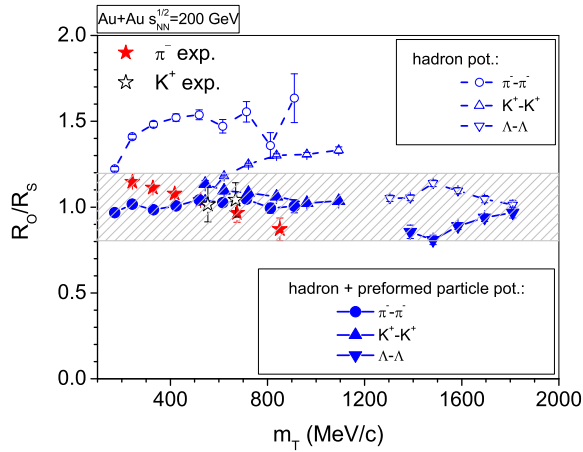}
\caption{$m_T$-dependence of the ratio $R_O/R_S$ of $\pi^-$, $K^+$,
and $\Lambda$ sources for ``s200'' case with potentials only for
formed baryons (lines with open symbols) and with potentials for
both formed and ''pre-formed'' particles (lines with solid symbols).
Solid and open stars represent the RHIC data of pions and kaons
separately \cite{Adler:2004rq,Adams:2004yc,Bekele:2004ci}. A $20\%$
deviation margin is marked in light gray.} \label{fig5}
\end{figure}

Fig.\ \ref{fig5} shows the $m_T$-dependence of the ratio $R_O/R_S$
of $\pi^-$, $K^+$, and $\Lambda$ sources at transverse momenta
$250<k_T<350$MeV$/c$ for central Au+Au reactions at
$\sqrt{s_{NN}}=200$ GeV. The calculations with potentials only for
formed baryons (lines with open symbols) and with potentials for
both formed and ''pre-formed'' particles (lines with solid symbols)
are compared with pion data \cite{Adler:2004rq,Adams:2004yc} and
preliminary kaon data \cite{Bekele:2004ci}. Considering the
uncertainties from the non-Gaussian effect \cite{Chung:2007yq} as
well as the variant treatments of Coulomb correction on FSI (see
\cite{Adamova:2002wi,Adams:2004yc}), a $20\%$ deviation margin is
marked in light gray. It is seen that, although the effect of the
formed hadron potentials on the ratio becomes weaker with the
increase of particle mass, the $R_O/R_S$ values of pions and kaons
at large $k_T$ are above the upper limit of the shadowed area. The
inclusion of ''pre-formed'' particle interactions cures these
deviations and allows for a consistent understanding of the data.
Here it should be addressed that due to the
$\widetilde{x}$-$\widetilde{t}$ correlation shown in
Eq.~(\ref{eqRo}), the $R_O/R_S \rightarrow 1$ does not mean that the
mean duration time should approach unity, a finite non-zero mean
proper duration time has been discovered by recent experiments
\cite{Afanasiev:2007kk} via three-dimensional source imaging methods
which definitely deserve further transport-model analysis.

To summarize, the $m_T$ dependence of the HBT parameters for pion,
kaon, and $\Lambda$ sources are investigated at three SPS and top
RHIC energies with a microscopic transport model (UrQMD). In the
cascade mode (i.e., without potential interactions), the HBT radii
$R_L$ and $R_S$ fall roughly with the $m_T$-scaling only at the
lowest SPS energy $E_{lab}=20A$ GeV, while the $R_O$ at all energies
and the $R_L$ at energies $E_{lab}=158A$ GeV and $\sqrt{s_{NN}}=200$
GeV are apart from the $m_T$ scaling line. Even the inclusion of
potentials for formed hadrons does not alter this picture. However,
with a density dependent potential for ``pre-formed'' particles from
string fragmentation, the HBT radii of pion and kaon sources, even
those of the $\Lambda$ source, are seen to follow the $m_T$-scaling
line $R=3/\sqrt{m_T}$ fairly well. The strong and Coulomb
interactions at late stage are observed to influence the HBT radii
and the $\lambda$ factor. The HBT ``t-puzzle'' in many calculations
exhibited by the larger $R_O/R_S$ ratios of pion and kaon sources
than data can only be solved by the ``pre-formed'' hadron potential
which dominates {\it at the early stage} of HICs.

\section*{Acknowledgements}
We would like to thank S. Pratt for providing the CRAB program and
acknowledge support by the Frankfurt Center for Scientific Computing
(CSC). We also thank S. Pratt, M. Lisa, Jan Kurcewicz, Zbigniew
Chajecki, Stephane Haussler for helpful discussions. Q. Li thanks
the Frankfurt Institute for Advanced Studies (FIAS) for financial
support. This work is partly supported by GSI, BMBF, and
Volkswagenstiftung.

%\newpage

\end{document}